\documentclass[aps,pre,twocolumn,float]{revtex4}
\usepackage{amsmath,bm,epsfig}
\newcommand{\B}[1]{{\bm{#1}}}
\usepackage[dvips]{color}
\begin{document}

\title{Magneto-mechanical Coupling in Thermal Amorphous Solids}
\author{H. George E. Hentschel$^{1,2}$, Valery Ilyin$^1$, Chandana Mondal$^1$, and Itamar Procaccia$^1$}
\affiliation{$^1$Department of Chemical Physics, The Weizmann Institute of
Science,  Rehovot 76100, Israel\\$^2$ Department of Physics,
Emory University, Atlanta, Georgia}
\begin{abstract}
Standard approaches to magneto-mechanical interactions in thermal magnetic crystalline solids involve Landau functionals in which the lattice anisotropy and the resulting magnetization easy axes are taken explicitly into account. In glassy systems one needs to develop a theory in which the amorphous structure precludes the existence of an easy axis, and in which the constituent particles are free to respond to their local amorphous surroundings and the resulting forces. We present a theory of all the mixed responses of an amorphous solids to mechanical strains and magnetic fields. Atomistic models are proposed in which we test the predictions
of magnetostriction for both bulk and nano-film amorphous samples. The application to nano-films
with emergent self-affine free interfaces requires a careful definition of the film ``width" and its
change due to the magnetostriction effect.
\end{abstract}
\maketitle
\section{Introduction}

\begin{figure}
     \vskip 1 cm
\includegraphics[scale = 0.32]{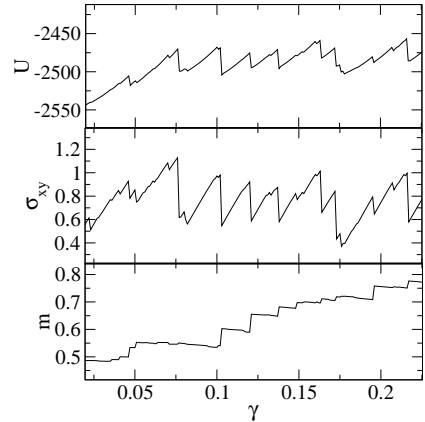}
\caption{Energy, stress and magnetization changes during plastic events as result
of increasing strain in a magnetic amorphous solids. The data is
taken from Ref.~\cite{16HPS} for an athermal example}.
\label{events}
\end{figure}

The subject of the interaction between mechanical and magnetic properties in magnetic glasses has been relatively neglected by theorists.
    Despite the enormous amount of work on magnetism in crystalline materials, (including ``spin glasses" where spins are restricted to reside on a lattice), and the equally enormous amount of work on non-magnetic glasses, there have been almost no theoretical studies of elastic, plastic and magnetic responses to shear and to external magnetic fields in glasses with magnetic properties until recently \cite{82Liv, 99DZ, 06Zap1, 12HIP,13DHPS,14HPS,14HIPS}. The crucial difference is that particles in a glass are free to move around whether they carry spins or not, and therefore
    there is a strong coupling between the mechanical and the magnetic properties of these materials. A generic plastic event in such materials is accompanied by simultaneous discontinuous change in stress, energy and magnetization cf. Fig.~\ref{events}.
A number of model glasses with magnetic interaction were put forward, allowing  highly accurate simulations for which one could offer
    detailed theories \cite{12HIP,13DHPS,14HPS,14HIPS}.  It was shown that magnetism can be induced by plastic events \cite{14HPS};  One could also study with exquisite detail the statistics of Barkhausen Noise in magnetic
    glasses to discover that it can belong to a number of different universality classes depending on the details of the magnetic interactions \cite{15DHJMPS}.

  One of the best known and important cross effects between mechanics and magnetism is magnetostriction (shape change due to to applied magnetic fields) and its inverse, the Villari effect (induced
  magnetization due to mechanical strain).
The theory of magnetostriction as applied to  solids and thin films typically make a number of implicit  assumptions  about the solids investigated. Often one presumes that the considered solid has a crystalline structure and the magnetic atoms lie at well defined lattice sites (up to thermal fluctuations). In consequence, due to spin-orbit coupling, there exist global magnetic easy axes along which the macroscopic magnetization prefers to orient at low temperatures. In the absence of such easy axes, if the only magnetic interaction is an exchange interaction, all directions are degenerate in the absence of an applied magnetic field.  Usually the assumption is also made that we are dealing with a low temperature situation in which the magnetization is saturated in magnitude and thus the easy axis controls the angle but not the magnitude of the magnetization $\bf m$. If these were the only magnetic energy terms there would not exist any magnetostriction. But there also exists strain energy in the solid and this is coupled to the  magnetization, resulting in a strained configuration of the solid as the lowest energy state of the system. This can be seen in the easiest way if we write down the Landau function $F$ for the magnetic and strain energies. Taking for example a cubic lattice at low temperatures $T\ll T_c$ with
saturated magnetization,
\begin{widetext}
\begin{eqnarray}
\label{landau}
F & = & K(\alpha_1^2\alpha_2^2+\alpha_2^2\alpha_3^2+\alpha_3^2\alpha_1^2)
   + B_1(\alpha_1^2\epsilon_{xx}+\alpha_2^2\epsilon_{yy}+\alpha_3^2\epsilon_{zz})
  + B_2(\alpha_1\alpha_2\epsilon_{xy}+\alpha_2\alpha_3\epsilon_{yz}+\alpha_3\alpha_1\epsilon_{zx})\nonumber \\
 &+& (1/2)c_{11}(\epsilon_{xx}^2+\epsilon_{yy}^2+\epsilon_{zz}^2)
 + (1/2)c_{44}(\epsilon_{xy}^2+\epsilon_{yz}^2+\epsilon_{zx}^2)
 + c_{12}(\epsilon_{yy}\epsilon_{zz}+\epsilon_{xx}\epsilon_{zz}+\epsilon_{xx}\epsilon_{yy}).
 \end{eqnarray}
 \end{widetext}
 Here $\alpha_i$ are the cosines of the angles between the magnetization direction and the cube easy axes $x_i$, while $\epsilon_{ij}$ is the strain tensor. $K$ is the strength of the anisotropy energy; $B_1$,$B_2$ are magnetoelastic coupling constants; while $c_{ij}$ are the elastic moduli of the cube. By minimizing the Landau functional with respect to strain $\partial F/\partial\epsilon_{ij}=0$ we can find explicit expressions for the strain in the material that minimizes the energy
\begin{eqnarray}
\label{strain}
\epsilon_{ii} &=& \frac{B_1[c_{12} - \alpha_i^2(c_{11}+2c_{12})]}{[(c_{11}-c_{12})(c_{11}+2c_{12})]} \nonumber \\
\epsilon_{ij} &=&\frac{ -B_2 \alpha_i \alpha_j}{c_{44}} .
 \end{eqnarray}
 Using Eq.~(\ref{strain}),  the magnetostriction in a direction ${\bf \beta} \equiv (\beta_1,\beta_2,\beta_3)$ where the $\beta_i$ are the cosines of the angles between the measurement direction and the cube axes $x_i$ is then given by
 \begin{equation}
 \label{ms}
 \delta \ell/\ell = \sum_{i\le j} \epsilon_{ij}\beta_i\beta_j.
 \end{equation}
This result was first derived by N. Akulov in 1926, \cite{26Aku}.

In this paper we focus on situations involving amorphous solids and metallic glasses, where the basic assumptions made in deriving Eqs.~(\ref{landau}) and~(\ref{strain}) do not apply as in amorphous solids there is typically no global easy axis in the material. Moreover, we are often interested in situations where we are in a glass phase $T<T_g$ but above or close to the  Curie temperature $T\approx T_c$ so that we cannot assume either that the magnetization is saturated or that applied magnetic fields $\bf B$ do not have a strong influence on the size of the magnetization $\bf m$. Finally Eq.~(\ref{landau}) is a macroscopic energy functional and we would like to look at magnetostriction in a more microscopic (atomistic) context in glasses.

The structure of this paper is as follows: in Sect.~\ref{micro} we present atomistic models
for magnetic glasses. In Sect.~\ref{genthe} we describe the general approach to the responses
of magnetic glasses to mechanical and magnetic strains. In Sect.~\ref{num} the numerical simulations
are presented, stressing results for magnetostriction in both bulk and film glasses. Section \ref{calc} applies the general results of Sect.~\ref{genthe} to extracting the magnetosriction coefficient in thermal glassy materials. Section \ref{summary} provides a summary of the paper and some discussion.

\section{Microscopic Models}
\label{micro}

The potential energy $U$ of $N$ point particles in an amorphous magnetic solids in the presence of a magnetic field $\B B$ can be written as
\begin{equation}
U (\{\B r_i\},\{{\B S}_i\})= U_{\rm mech}(\{\B r_i\}) + U_{\rm mag}(\{{\B r}_i\}, \{{\B S}_i\}; \B B) \ ,\label{Ham}
\end{equation}
where $\{\B r_i\}_{i=1}^N$ are the positions of the particles and $\B S_i$ are spin variables.
\subsection{The Mechanical Interactions}
\label{mechint}

The mechanical part of our Hamiltonian Eq.~(\ref{Ham}) can be taken as any of the standard models of glass formers, and we will
assume that it is a sum of binary interactions such that
\begin{equation}
\label{umech}
U_{mech}(\{\B r_i\}) = \sum_{<ij>} \phi(r_{ij}),
\end{equation}
where $\langle ij \rangle$ means "all distinct pairs", $r_{ij} \equiv |\B r_i-\B r_j|$ are the instantaneous distances between particles $i$ and $j$.
This still leaves a lot of freedom, as the literature attests to a variety of models with
binary interactions that produce good glass formers.
To generate a glass, we simulate Kob-Andersen  binary mixture of two types of  particles \cite{KA1,KA2,KA3}.
We address the particles  as type-A particles which are magnetic and the other type of particles, which are non-magnetic, as type-B particles. The ratio of number of $A$ and $B$ type particle is taken as
$80:20$. The mechanical part of interatomic interactions are defined  by truncated and shifted  Lennard-Jones
potentials
\begin{equation}
\phi_{ij}(r)=\left\{
\begin{array}{ll}
\phi_{ij}^{LJ}(r)+C_{ij}
&\textrm{if $r\le R^{cut}_{ij}$,}\\
0&\textrm{if $r> R^{cut}_{ij}$,}
\end{array}\right.
\label{KA1}
\end{equation}
where $C_{ij}=-\phi_{ij}^{LJ}(R^{cut}_{ij})$ and
\begin{equation}
\phi_{ij}^{LJ}(r)=4e_{ij}
\bigg[\bigg(\frac{\sigma_{ij}}{r}\bigg)^{12}
-\bigg(\frac{\sigma_{ij}}{r}\bigg)^{6}\bigg].
\label{LJ}
\end{equation}
To simplify the simulations the pair interactions in Eq.~(\ref{KA1}) are truncated at distance $R^{cut}_{ij}=2.5\sigma_{ij}$.
It is convenient to introduce reduced units, with $\sigma_{AA}$ being the unit of length and
$e_{AA}$ the unit of energy. Parameters for $A-B$ and $B-B$ interactions are given by
$\sigma_{BB}/\sigma_{AA}=0.88$, $\sigma_{AB}/\sigma_{AA}=0.8$, $e_{BB}/e_{AA}=0.5$ and $e_{AB}/e_{AA}=1.5$. The reported glass transition temperature $T_g$ of the Kob-Andersen binary mixture in $3D$ ~\cite{KA3} is $T_g=0.28$.

\subsection{The Magnetic Interactions}
\label{choice}
The magnetic properties of amorphous magnets are extremely varied and cannot be represented by a unique Hamiltonian. For example spins can be effectively localized on individual atoms, or be dominated by delocalized spins on conduction electrons. If localized the total angular momentum of unpaired electrons will depend on the atomic species considered. In this paper we shall consider the simplest Heisenberg magnetic Hamiltonian that couples magnetic and mechanical properties in amorphous solids -- namely an exchange interaction in which the exchange integral is an explicit function of particle positions
\begin{equation}
\label{umag}
U_{mag}(\{{\B r}_i\}, \{{\B S}_i\};\B B) =  - \sum_{<ij>}J(r_{ij}){\bf S}_i \cdot {\bf S}_j  -g\mu_B\sum_i {\bf S}_i\cdot {\bf B} \ .
\end{equation}
The first term on the right hand side is the short range exchange interaction $U_{\rm ex}(\{{\B r}_i\}, \{{\B S}_i\})$. Typically  the exchange integral $J(r)>0$, thus encouraging ferromagnetism, and will be peaked at some distance $r_1$ or else be an exponentially decreasing function of $r$. Note that $ J(r_{ij})$ will couple magnetism and strain in a nontrivial fashion. The exchange energy of interaction  between Heisenberg spins is chosen, following Ref.~\cite{LWABS94} as a Yukawa type potential of the form
\begin{equation}
J(x)=J_0 \frac{\exp(-\kappa x)}{x}.
\label{Yuk}
\end{equation}
Like the Lennard-Jones interaction, the exchange interaction is also truncated at $x=2.5$ and shifted
to zero at that point. The screening parameter
$\kappa$ determines the range of the interaction. We have taken $\kappa=3.6$. Finally, in our case $J_0=3.0$.

\section{Mechanical and Magnetic Responses at Finite Temperatures}
\label{genthe}
The theory of mechanical and magnetic responses of amorphous solids at zero temperature
is available, and for completeness we summarize the main results in Appendix \ref{zero}.
Here we present the theory for thermal glasses, taking into account the effects
of thermal fluctuations.
Given any dynamical variable $Y(\{\B r_i,\B S_i\}_{i=1}^N)$ its thermal average is
determined by
\begin{eqnarray}
\langle Y \rangle&=&\frac{\int d X d S~ Y e^{-\beta U(X,S)}}{Z}\nonumber\\
Z&=& \int d X d S~ e^{-\beta U(X,S)}\nonumber\\
X&\equiv& \{\B r_i\}_{i=1}^N\ ,\quad S\equiv \{\B S_i\}_{i=1}^N \ .
\label{defs}
\end{eqnarray}
Noticing that $\langle Y \rangle$ is a function of the magnetic field and the mechanical strain we can compute
\begin{eqnarray}
\frac{\partial \langle Y \rangle}{\partial B_\alpha} &=&\Big\langle \frac{\partial  Y }{\partial B_\alpha}\Big\rangle -\beta\left[\Big \langle Y\frac{\partial U}{\partial B_\alpha}\Big\rangle-\Big\langle
\frac{\partial U}{\partial B_\alpha}\Big\rangle\Big\langle Y\Big\rangle\right ]\ , \nonumber\\
\frac{\partial \langle Y \rangle}{\partial \epsilon_{\alpha\beta}} &=&\Big\langle \frac{\partial  Y }{\partial \epsilon_{\alpha\beta}}\Big\rangle -\beta\left[\Big \langle Y\frac{\partial U}{\partial \epsilon_{\alpha\beta}}\Big\rangle-\Big\langle
\frac{\partial U}{\partial \epsilon_{\alpha\beta}}\Big\rangle\Big\langle Y\Big\rangle\right ]\ .
\label{responses}
\end{eqnarray}
Specializing these equation to the stress tensor and the magnetization we get the explicit
expressions for the magnetostriction and the Villari effects in thermal systems:
\begin{eqnarray}
\label{ms1}
\!\frac{\partial \langle \sigma_{\alpha\beta} \rangle}{\partial B_\gamma}\!\! &=&\!\!\Big\langle \frac{\partial  \sigma_{\alpha\beta} }{\partial B_\gamma}\Big\rangle \!-\!\beta\!\left[\Big \langle \sigma_{\alpha\beta}\frac{\partial U}{\partial B_\gamma}\Big\rangle\!-\!\Big\langle
\frac{\partial U}{\partial B_\gamma}\Big\rangle\Big\langle \sigma_{\alpha\beta}\Big\rangle\right ]\ , \\
\!\!\!\frac{\partial \langle m_\gamma\rangle}{\partial \epsilon_{\alpha\beta}}\!\! &=&\!\!\Big\langle \frac{\partial  m_\gamma }{\partial \epsilon_{\alpha\beta}}\Big\rangle \!-\!\beta\left[\Big \langle  m_\gamma\frac{\partial U}{\partial \epsilon_{\alpha\beta}}\Big\rangle\!-\!\Big\langle
\frac{\partial U}{\partial \epsilon_{\alpha\beta}}\Big\rangle\Big\langle  m_\gamma\Big\rangle\right ]\ . \label{vil}
\end{eqnarray}
Magnetostriction is actually determined by the strain response to the magnetization, but using
linear elasticity theory we can easily invert from the stress response to the strain response,as
\begin{equation}
\sigma_{\alpha\beta} = c_{\alpha\beta\gamma\delta} \epsilon_{\gamma\delta} \ .
\end{equation}
where $\B c $ is the elastic modulus tensor. Eqs.~(\ref{ms1}) and (\ref{vil}) can be simplified further by using the identities
\begin{eqnarray}
 \frac{\partial U}{\partial B_\gamma} &=& -V n m_\gamma \ , \quad m_\gamma = \frac{g\mu_B}{N_s}  \sum \B S_i\\
 \frac{\partial U}{\partial \epsilon_{\alpha\beta}}&=& V\sigma_{\alpha\beta} \ .
\end{eqnarray}
Here $n$ is the number of spins per unit volume and $N_s$ the number of particles carrying spins
in the system. The reader should note that here $m_\gamma$ and
$\sigma_{\alpha\beta}$ are the instantaneous variable rather than thermal averages. Using these identities in  Eqs.~(\ref{ms1}) and (\ref{vil}) we get the final results
\begin{eqnarray}
\label{ms2}
\!\frac{\partial \langle \sigma_{\alpha\beta} \rangle}{\partial B_\gamma}\!\! &=&\!\!\Big\langle \frac{\partial  \sigma_{\alpha\beta} }{\partial B_\gamma}\Big\rangle \!+\!\beta V n \left[\Big \langle \sigma_{\alpha\beta}m_\gamma\Big\rangle\!-\!\Big\langle
m_\gamma\Big\rangle\Big\langle \sigma_{\alpha\beta}\Big\rangle\right ]\ , \\
\!\!\!\frac{\partial \langle m_\gamma\rangle}{\partial \epsilon_{\alpha\beta}}\!\! &=&\!\!\Big\langle \frac{\partial  m_\gamma }{\partial \epsilon_{\alpha\beta}}\Big\rangle \!-\!\beta V \left[\Big \langle  m_\gamma \sigma_{\alpha\beta}\Big\rangle\!-\!\Big\langle
 \sigma_{\alpha\beta}\Big\rangle\Big\langle  m_\gamma\Big\rangle\right ]\ . \label{vil2}
\end{eqnarray}
From the identity of the mixed second derivatives of $U$ we can derive immediately the
Maxwell relation
\begin{equation}
\frac{\partial \langle \sigma_{\alpha\beta} \rangle}{\partial B_\gamma}= - n \frac{\partial \langle m_\gamma\rangle}{\partial \epsilon_{\alpha\beta}} \ .
\end{equation}
In the next subsection we motivate further discussion by specializing to Lennard-Jones glass formers.
\section{Numerical Simulations}
\label{num}
The numerical creation of glassy bulk and film phases and their equilibration using Monte Carlo techniques is presented in Appendix \ref{A}. We discuss separately the results for bulk and film.
\subsection{Bulk phase}
In the bulk phase, we simulate $N=4000$ particles contained in a cell endowed with periodic boundary condition in all three directions to mimic an infinite system. The pressure and the temperature were  fixed at $P=2.2$  and  $T=0.23$ (below $T_g$ but above $T_c$).

\begin{figure}[!h]
\centering
\epsfig{width=.38\textwidth,file=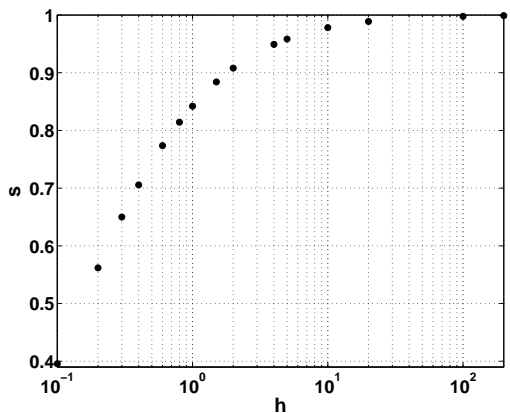}
\caption{Dependence of $s$ on the applied magnetic field for the bulk magnetic glass.
Here $h\equiv g\mu_B B$}
\label{fig1}
\end{figure}
\begin{figure}[!h]
\centering
\epsfig{width=.38\textwidth,file=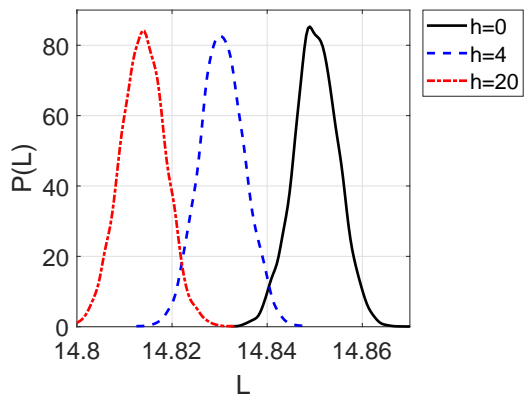}
\caption{$P(L)$ in NPT simulations of the amorphous glass at different values of the applied magnetic field.}
\label{fig2}
\end{figure}
Taking the external field to point in the $z$ direction, we define
\begin{equation}
s \equiv \Big\langle \frac{\sum_i S^z_i}{N_s} \Big\rangle \ .
\end{equation}
This quantity was computed in an NPT ensemble and its dependence on the external field is shown in Fig.~\ref{fig1}. At zero field the system is disordered and at high values of the field  the magnetization saturates.
In Monte Carlo simulations at fixed pressure the average volume changes with increasing the external field. This is measured by the changing length $L$ of the simulation cell; the probability distribution function (pdf) $P(L)$ at different values of applied external field $h$ is shown in Fig.~\ref{fig2}. In general, this magnetostriction effect is weak but clearly observable.
\begin{figure}
\centering
\epsfig{width=.38\textwidth,file=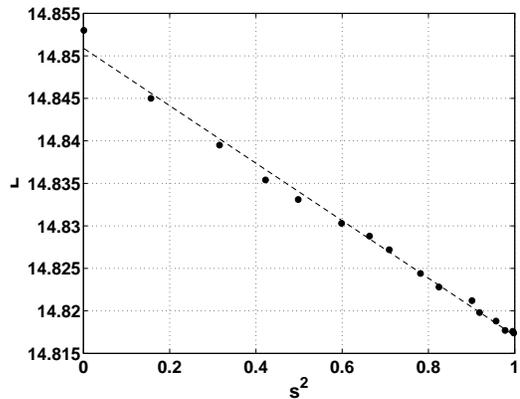}
\caption{Dependence  of average value of $L$ in NPT simulations $s^2$}
\label{fig3}
\end{figure}
The magnetostriction is quantified as the fractional change in length of the sample $\gamma=\delta L / L$. When $\gamma$ is measured at high values of the external magnetic field (i.e. when the
magnetization saturates) one refers to the ``saturation magnetostriction".
In fact it is advantageous to define the magnetostriction effect by its dependence on $s$ (see, e.g., \cite{LG06}). Indeed, the experimental results in Ref.~\cite{LG06} exhibit a linear
 dependence of the magnetostriction when plotted as a function of $s^2$.
\begin{figure}
\centering
\epsfig{width=.38\textwidth,file=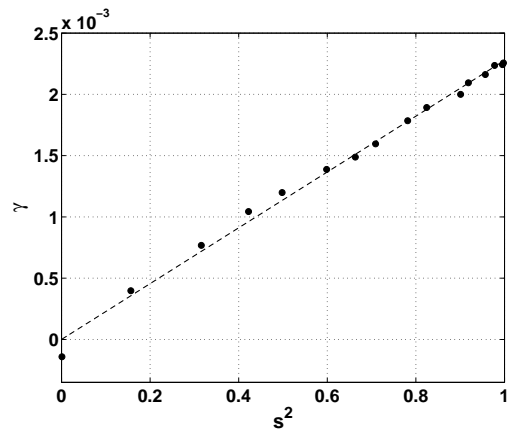}
\caption{Magnetostriction in bulk phase.}
\label{fig4}
\end{figure}
The same result was obtained analytically using the model discussed in \cite{CG09}.
Results of our simulations for the bulk phase are in agreement, cf. Fig.~\ref{fig3} where the dependence of $L$ on $s^2$ is observed. Fitting the data by least squares and denoting the length of the simulation cell at zero magnetization $L_0$ the quantity $\gamma=(L_0-L)/L_0$ was calculated. The dependence of $\gamma$ on $s^2$ is shown in Fig.~\ref{fig4}. These results are in agreement with Eq.~(\ref{mstmag}).
Therefore, one can write
\begin{equation}
\gamma=\lambda s^2 \ .
\label{mstmag}
\end{equation}

\subsection{Film Construction}
A nano-thin film is generated on top of a face-centered cubic crystalline substrate composed of fixed $N_S=1152$ identical particles interacting via Lennard-Jones potential with the
film particles. The interaction parameters are : $\sigma_{SA}/\sigma_{AA}=\sigma_{SB}/\sigma_{AA}=0.8$,
$e_{SA}/\epsilon_{AA}=e_{SB}/e_{AA}=1.5$. The subscript $S$ stands for substrate.
\begin{figure}[!h]
\centering
\epsfig{width=.38\textwidth,file=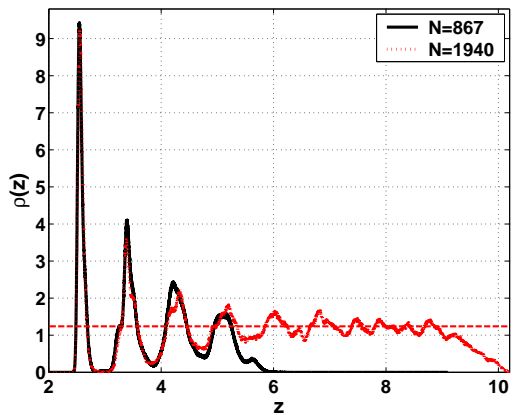}
\caption{The local density as a function of $z$ for two films containing different number of particles, $N=867$ and $N=1940$ respectively. The red constant line represents the average density of the bulk phase from which the film was created.}
\label{fig5}
\end{figure}
The substrate particles are taken to be non-magnetic.  The substrate density in equilibrium is $\rho_S=2.1$, providing a support to the film whose lateral dimensions are $L\times L$. To create a film of the binary mixture we return to our bulk simulation and cut off
a slab of desired width $w$ perpendicular to the $z$ direction with lateral dimensions $L\times L$.  The geometric fit of the film slab to the substrate is obtained by taking the bulk
at density $\rho=1.22$.

When the slab is positioned on the substrate it is placed with a gap between the film and the substrate which is $2^{1/6}\sigma_{SA}$ at the minimum of the Lennard-Jones potential between the substrate and the A particles. The created simulation box is kept periodic in the $x$ and $y$ directions with the length of the periodicity cell begin $L=14.86$.  The substrate acts as a fixed wall at the bottom of the film. In order to create an equilibrated film we first impose a maximal extent of the film
on the upper boundary.  Switching on the Monte Carlo algorithm, when a particle attempts to cross the
upper boundary, this move is rejected. The initial gap between the upper particles in the film and and this boundary is $2.5\sigma_{AA}$. Clearly, when the NVT Monte Carlo steps accumulate, the film adjusts the height along $z$ creating an upper free boundary.

\subsection{Computing the film width}
\begin{figure}[!h]
\centering
\epsfig{width=.38\textwidth,file=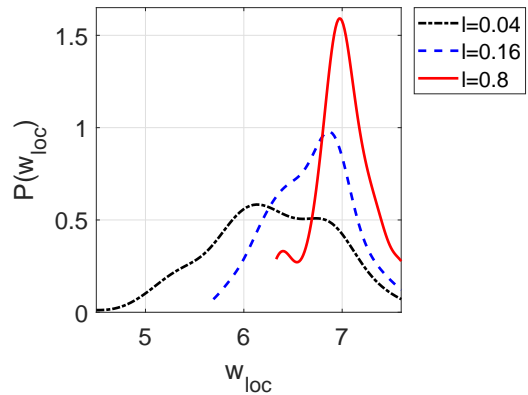}
\caption{Example of distributions of the local width for different size of subdomains $l\times l$ in a film at temperature $T=0.23$.}
\label{fig6}
\end{figure}

Due to the existence of the substrate at the bottom ($z=0$ is defined at the top of the crystalline substrate) and a free film at the top, translational invariance in the $z$ direction is lost. Indeed, the local density along $z$ for two films of different widths is shown in Fig.~\ref{fig5}.
Like in liquid phases (see \cite{AD14}) the films show ordering near the substrate. In the wider film the region farther from the substrate reaches the average density in the bulk phase. In both cases near the free boundary there is a smooth cross-over from the dense phase to vacuum. This makes the influence of film topography on the film width non-trivial. For large films the surface will be self-affine.
Here we are specifically interested in how to treat nano-films in which the fluctuations of the
free surface are of the order of the film width.

In general the width of a given film will depend on the number of particles $N$ and
the magnetization $s$.
In order to define the film width, the area of the simulation cell is divided into sub-domains of size $l\times l$.
\begin{figure}
\centering
\epsfig{width=.38\textwidth,file=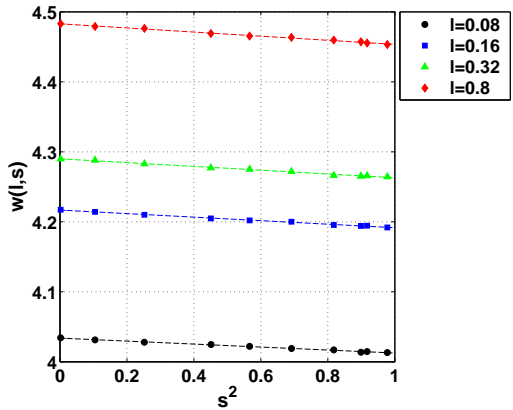}
\caption{Dependence of the film width on the magnetization at different values of $l$.  Particle number in the simulation cell $N=1940$.}
\label{fig7}
\end{figure}
In each sub-domain, one creates a list of  particles ordered according to their $z$-coordinate from top to bottom. After every Monte-Carlo sweep we determine again this ordering.  In each realization the  particle with highest value of $z$ in each sub-domain determines the position of the free surface, defined as $w_{loc}$. The typical distributions of the local widths in a film of binary mixture for different $l$ is shown in Fig.~\ref{fig6}. The position of the maximum of this distribution defines  ``the width"  of the film $w(l,s)$. As one can see the estimated width of a film increases with increasing $l$.
The dependence of the width on $s$ measured at different values of $l$ is shown in Fig.~\ref{fig7}. Examining Fig.~\ref{fig7} we conclude that the apparent width $w(l,s)$ depends
linearly on $s^2$ with a different intercept and slope for each $l$. These functions can be summarized
\begin{figure}
\centering
\epsfig{width=.38\textwidth,file=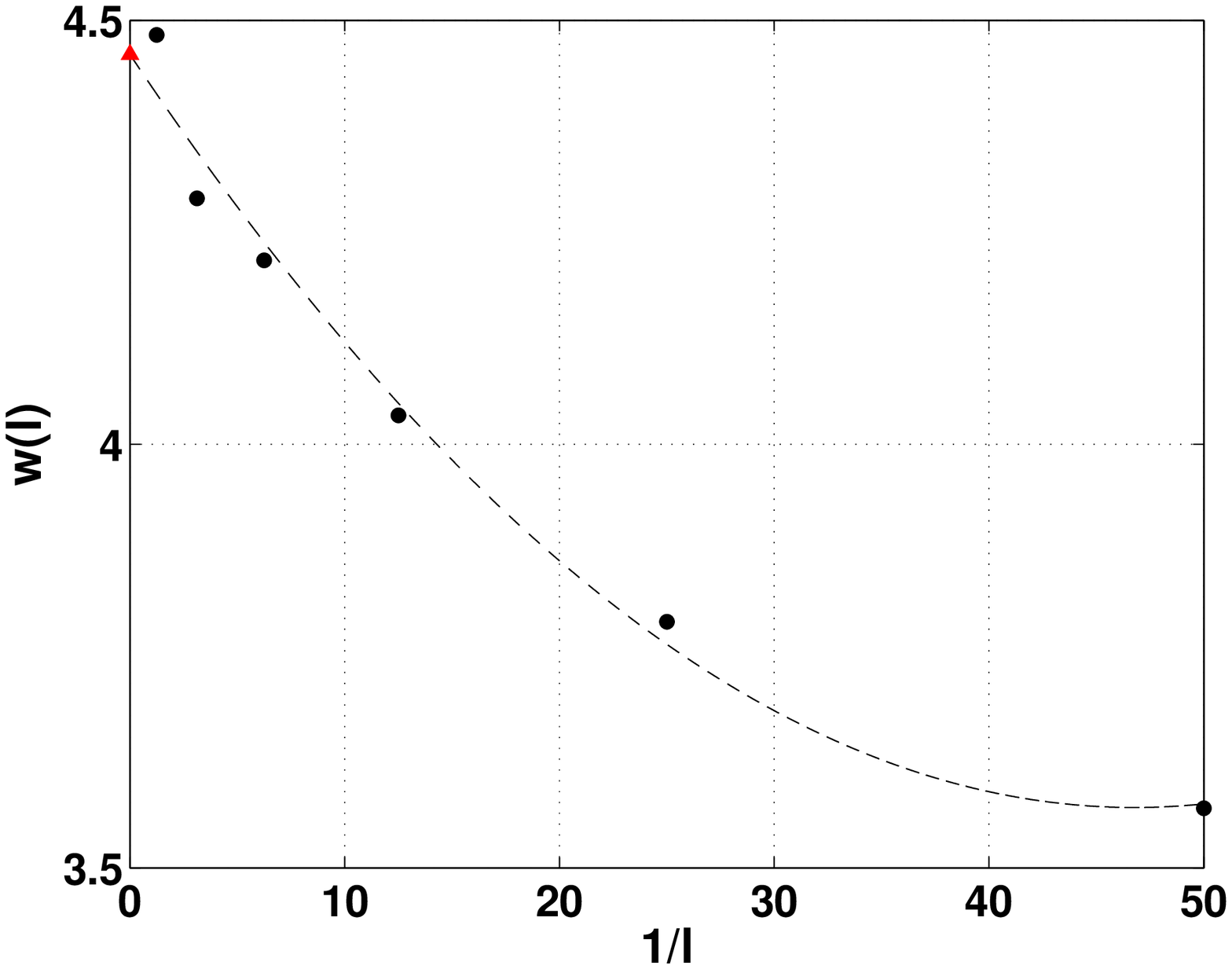}
\epsfig{width=.38\textwidth,file=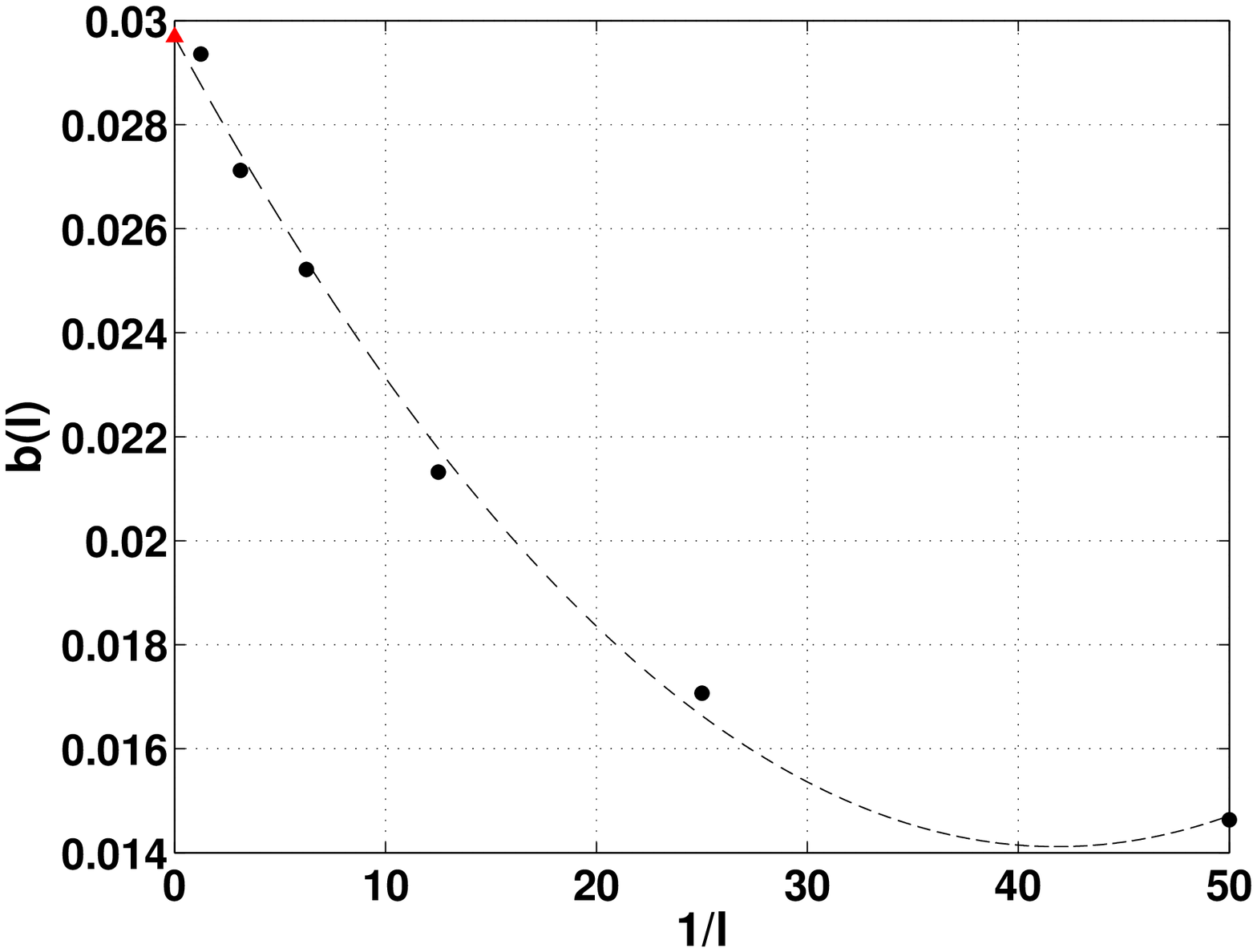}
\caption{Top panel - dependence of the coefficient $w(l)$ on the inverse value of $l$.  Extrapolating to $l\to \infty$ we estimate the film width in the absence of the external magnetic field $w(l\to \infty)=w_0$ The red triangle indicates extrapolated value $w_0= 4.46$. Bottom panel -  dependence of the coefficient $b(l)$ on the inverse value of $l$. Extrapolated to $l\to \infty$ value of this coefficient is related to the magnetostriction coefficient of the film by $b=b(l\to \infty)=\lambda w_0$. The red triangle indicates extrapolated value $b= 0.03$. Coefficients in Eq.~(\ref{linw}) are estimated from the data shown in Fig.~\ref{fig7} for the film with  particle number in the simulation cell $N=1940$.}
\label{fig8}
\end{figure}
by the equation
\begin{equation}
w(l,s) =w(l)-b(l) s^2 \ ,
\label{linw}
\end{equation}
where $w(l)\equiv w(l,s=0)$.
Careful fitting indicates that $b(l)$ has a systematic dependence on $l$.  From this equation
we can extract the $l$ dependent magnetostriction coefficient as
\begin{equation}
\gamma(l) =\frac{w(l)-w(l,s)}{w(l)} =\lambda(l) s^2 ,
\label{msell}
\end{equation}
where $\lambda(l)=b(l)/w(l)$.
Finally, the magnetostriction coefficient for the films is defined as
\begin{equation}
\gamma=\gamma(l\to \infty)  =\lambda s^2 \ ,
\label{deflam}
\end{equation}
where  $\lambda \equiv \lambda(l \to \infty )$. Repeating the procedure in films
with different number of particles $N$ we find that $\lambda$ has a dependence on $N$
and we attempt next to determine the ``best" value of $\lambda$.
To this aim we fit the data in Fig.~\ref{fig7} and extrapolate the slopes and intercepts
to $l\to \infty$. This is done in Fig.~\ref{fig8}. It follows from Eq.~(\ref{linw}), Eq.~(\ref{msell}) and Eq.~(\ref{deflam}) that for a given film with $N$ particles
the film width at $s=0$ is defined by $w(l\to\infty)=w_0(N)$ and the slope in Eq.~(\ref{linw}) tends to $b(l\to\infty)=\lambda w_0(N)$.
\begin{figure}
\centering
\epsfig{width=.38\textwidth,file=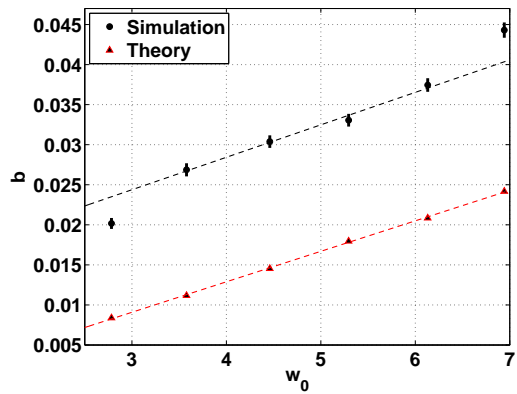}
\caption{The coefficient $b$ as a function of the film width $w_0$. Simulational results are
respersented by black dots.  The theoretical results are the mean-field theory presented in Sec.~\ref{MS} culminating with Eq.~(\ref{magneto})}
\label{fig9}
\end{figure}

 In Fig.~\ref{fig9} we plot simulations result for $b \equiv b(l\to\infty)$ as a function of $w_0$ as the black circles.
 The linear character of this relationship indicates that the saturation
 magnetostriction of a wide enough films depends on their width as (see also \cite{SGS08})
\begin{equation}
\lambda=\lambda_{\rm bulk}+\frac{\Lambda}{w_0} \ ,
\label{filmlamb}
\end{equation}
where $\lambda_{\rm bulk}$ is the bulk value which is obtained when the width of the film tends
to infinity.
 For
 purposes here we are interested in the slope of $b$ vs $w_0$ ($\lambda_{\rm bulk}$), not in the intercept (we will discuss the intercept
 below). Fitting a least-square line to the data we estimate
 \begin{equation}
 \lambda_{\rm bulk} \approx (4.5\pm 0.2)\times 10^{-3} \ .
 \label{numres}
 \end{equation}
 In the rest of this paper we provide the theory which will culminate in a first principle
 evaluation of this magnetosriction coefficient and we will also argue how we expect
 it to change with the width of the film.

\section{Calculation of the magnetostriction}
\label{calc}

In this section we develop the general theory described in Sec.~\ref{genthe} with the aim
of rationalizing the results obtained in the numerical simulations. Our aim is to compute
$\langle \epsilon_{zz}\rangle$ as a function of $s$. As usual, since the strain
tensor $\B \epsilon$ is not a state variable, we need to compute the stress response $\B \sigma$
and extract the strain tensor from standard relations of elasticity theory. Thus our starting point
is Eq.~(\ref{ms2}) for the response of the stress to the magnetic field. The RHS of this equations
contains two terms, the Born term and the non-affine fluctuations term.
\subsection{The non-affine term}
To gain insight on the non-affine term in Eq.~(\ref{ms2}) we employ our simulations. Measuring
 the two terms at temperature $T=0.23$ we
conclude that the two terms cancel each other and the stress and spin fluctuations
 are practically decoupled. In fact, dividing the difference between the two terms by the magnitude
 of either of them we find numbers of the order $10^{-5}-10^{-4}$ for all values of $m$. The same result was obtained for all the
 non affine components in the response of the tensor $\langle \B \sigma \rangle$. We thus
 conclude that to a high approximation
 \begin{equation}
 \Big \langle \sigma_{\alpha\beta}m_\gamma\Big\rangle\!\approx \!\Big\langle
m_\gamma\Big\rangle\Big\langle \sigma_{\alpha\beta}\Big\rangle \ .
\label{MF}
 \end{equation}
 It should be stated that in the ferromagnetic
 phase at low temperatures this conclusion may change drastically, and see for example
 the $T=0$ results in Appendix \ref{zero}. Indeed, the athermal results in Refs.~\cite{12HIP,13DHPS,14HPS,14HIPS}
 show that non-affine contributions to the magnetostriction and other responses are
 comparable in size their respective Born terms. This can also be seen in the results
 quoted in Fig.~\ref{events} where the discontinuities in all the measured quantities are
 due to non-affine responses.

 A corollary of Eq.~(\ref{MF}) is that in calculating the Born term in the response
 of the stress tensor we can employ a mean-field decoupling between the stress and
 the magnetization fluctuations. This simplifies the analytic calculation considerably.

\subsection{Calculation of the Born term for magnetostriction}
\subsubsection{Definition of the strain tensor}
In thermal systems the particles are restricted dynamically to their cages for long enough time $\tau$ which is nevertheless shorter than the diffusion time. Therefore we can compute
a temperature dependent average position and an average spin orientation:
\begin{equation}
\langle \B r_i \rangle =\frac{1}{\tau} \int_0^\tau dt ~\B r_i(t)\ , \quad \langle \B S_i \rangle =
\frac{1}{\tau} \int_0^\tau dt ~\B S_i(t) \ .
\end{equation}
Once we apply an external strain $h_{ij}(\gamma)$ and a magnetic field $\B B$ the average position and
average spin orientation will experience an affine and a non-affine response.
\begin{equation}
\langle \B r_i \rangle (\gamma,B)=h_{ij}(\gamma) \langle \B r_j\rangle + \B u_i(\gamma, B) \ .
\end{equation}
Here $\B u_i$ is the non-affine response that takes place as a result of the affine
external strain and magnetic field, after which the system returns to thermal equilibrium.
Defining the strain
tensor $\epsilon_{\alpha\beta}$ in terms of the change in distance between pairs of particles (cf. for example \cite{LL})
\begin{equation}
\langle r_{ij} \rangle (\gamma,B) \equiv \langle r_{ij} \rangle  \sqrt{1 +2 \frac{\epsilon_{\alpha\beta}\langle r^\alpha_{ij}\rangle\langle r^\beta_{ij}\rangle}{\langle r_{ij} \rangle^2}}  \ ,
\end{equation}
where $r^\alpha_{ij}$ is the $\alpha$ component of $\B r_{ij}$.
Expanding this transformation to second order
in $\epsilon_{\alpha\beta}$ we find,
\begin{eqnarray}
\label{rijstrain}
&&\langle r_{ij} \rangle(\gamma,B)= \langle r_{ij}\rangle +  \epsilon_{\alpha\beta}\frac{\langle r^\alpha_{ij}\rangle\langle r^\beta_{ij}\rangle}{\langle r_{ij}\rangle} \nonumber\\&&-\frac{\epsilon_{\alpha\beta} \epsilon_{\gamma\delta}\langle r_{ij}^\alpha \rangle \langle r_{ij}^\beta \rangle \langle r_{ij}^\gamma\rangle\langle r_{ij}^\delta\rangle}{2\langle r_{ij}\rangle^3 }\ .
\end{eqnarray}
 For the analysis below we define
\begin{equation}
\label{drijstrain}
\delta r_{ij}  \equiv  \langle r_{ij} \rangle(\gamma,B)- \langle r_{ij}\rangle  \ .
\end{equation}
\subsubsection{The affine stress response}
 To compute $\langle \partial \sigma_{\alpha\beta}/\partial B_\gamma\rangle$ we express it
in the form
\begin{eqnarray}
\Big\langle \frac{\partial \sigma_{\alpha\beta}}{\partial B_\gamma}\Big \rangle = \frac{1}{V}\Big\langle \frac{\partial^2 U}{\partial \epsilon_{\alpha\beta}\partial B_\gamma}\Big\rangle \ .
\end{eqnarray}
Under a strain $\epsilon_{\alpha\beta}$ the mechanical energy (Lennard Jones interaction) will transform as:
\begin{equation}
\label{Umechstrain1}
U_{mech}=\sum\limits_{i \neq j}\phi(r_{ij}) + \sum\limits_{i \neq j}\frac{d\phi}{dr_{ij}} \delta r_{ij} + \frac{1}{2}\sum\limits_{i \neq j}\frac{d^2\phi}{dr^2_{ij}} \delta r^2_{ij}
\end{equation}
Using Eq.~(\ref{drijstrain}) we can now write
\begin{equation}
\label{Umechstrain3}
U_{mech} = U_{mech}(\gamma=0)  + V\bigg[\epsilon_{\alpha\beta} c_{\alpha\beta}^{(1)}  + \frac{1}{2}\epsilon_{\alpha\beta}\epsilon_{\gamma\delta} c_{\alpha\beta\gamma\delta}^{(2)}\bigg] ,
\end{equation}
where
\begin{eqnarray}
\label{Umechstrain4}
c_{\alpha\beta}^{(1)}&=& (1/V) \sum\limits_{i \neq j}\frac{d\phi}{dr_{ij}}\frac{r^\alpha_{ij}r^\beta_{ij}}{r_{ij}} \nonumber \\
c_{\alpha\beta\gamma\delta}^{(2)}\!\!&=& \!\!(1/V) \sum\limits_{i \neq j}\!\!
\Bigg[\frac{1}{r^2_{ij}} \frac{d^2 \phi}{d r^2_{ij}}-\frac{1}{r^3_{ij}}\frac{d\phi}{d r_{ij}}\Bigg] r_{ij}^\alpha r_{ij}^\beta r_{ij}^\gamma r_{ij}^\delta .
\end{eqnarray}
\noindent We can now express the mechanical contribution to the stress tensor $\sigma_{\alpha\beta}^{mech} = \partial \langle U_{mech} \rangle/V\partial \epsilon_{\alpha\beta}$ as
\begin{equation}
\label{Umechstrain3}
\sigma_{\alpha\beta}  = c_{\alpha\beta}^{(1)}  + \epsilon_{\gamma\delta} c_{\alpha\beta\gamma\delta}^{(2)} \quad \text{without exchange interaction}\ .
\end{equation}
\noindent This is the stress in the material in the paramagnetic state in the absence of external magnetic fields. Let us now  consider the additional strain imposed on the material in the ferromagnetic state or due to the application of an external magnetic field.

\subsubsection{Exchange Energy Under Strain in the Glass Phase}

\noindent Under a strain $\epsilon_{\alpha\beta}$ the exchange coefficient $J(r_{ij})$ transforms as
\begin{equation}
\label{Jstrain}
J(\langle r_{ij}\rangle(\gamma)) = J(\langle r_{ij}\rangle) + \frac{dJ(r_{ij})}{dr_{ij}}\epsilon_{\alpha\beta}\frac{\langle r_{ij}^\alpha\rangle  \langle r_{ij}^\beta\rangle }{\langle r_{ij}\rangle} + O(\epsilon^2) \ ,
\end{equation}
where derivatives $dJ(x)/dx$ are always computed at $x=\langle r_{ij}\rangle$. We have only expanded $J(r_{ij})$ to first order in the strain as in the exchange energy this will generate a term of $O(\epsilon s^2)$ and therefore the second order term can be neglected. The effect of strain $\epsilon_{\alpha\beta}$ on the  average exchange interaction in the glass phase  is then given by $\langle U_{ex}\rangle  = -\sum\limits_{i \neq j}J(r_{ij}) \langle \mathbf S_i\cdot\mathbf S_j \rangle $. To estimate this term let us make the mean field approximation  and replace  $\langle \mathbf S_i\cdot\mathbf S_j \rangle \approx <S_i><S_j> = s^2$. Then we expand this energy to first order in  $\epsilon_{\alpha\beta}$ as
\begin{equation}
\label{Uexstrain}
 <U_{\rm ex}>  = - s^2[ \sum\limits_{i \neq j}J(\langle r_{ij}\rangle )
+ \frac{dJ(r_{ij})}{dr_{ij}} \epsilon_{\alpha\beta} \frac{\langle r_{ij}^\alpha\rangle  \langle r_{ij}^\beta\rangle }{\langle r_{ij}\rangle }] \ ,
 \end{equation}
or
\begin{equation}
\label{Uexstrain2}
  <U_{\rm ex}>= - V\bigg(a+\epsilon_{\alpha\beta}b_{\alpha\beta}\bigg)s^2
 \end{equation}
 where we have used the notation
 \begin{eqnarray}
 \label{Uexstrain3}
 a & = &(1/V)\sum\limits_{i \neq j}J(\langle r_{ij}\rangle ) \nonumber \\
b_{\alpha\beta}&=&(1/V)\sum\limits_{i \neq j}\frac{dJ(r_{ij})}{dr_{ij}}\frac{r^\alpha_{ij}r^\beta_{ij}}{\langle r_{ij}\rangle }
\end{eqnarray}
Combining the effects of strain on the mechanical and magnetic energies we find
\begin{eqnarray}
\label{Ustrain5}
&&\langle U\rangle =\langle U_{\rm mech}+U_{\rm ex} \rangle
=U_{\rm mech}(\gamma=0)\nonumber  \\&&+ V\bigg[\epsilon_{\alpha\beta} c_{\alpha\beta}^{(1)}  + \frac{1}{2}\epsilon_{\alpha\beta}\epsilon_{\gamma\delta} c_{\alpha\beta\gamma\delta}^{(2)}-\bigg(a+\epsilon_{\gamma\delta}b_{\gamma\delta}\bigg)s^2 \bigg] \nonumber \\
\end{eqnarray}
So
\begin{equation}
\label{Ustrain6}
\sigma_{\alpha\beta}(m) =  \Big\langle \frac{\partial U}{V \partial \epsilon_{\alpha\beta}}\Big\rangle = c_{\alpha\beta}^{(1)}+ \epsilon_{\gamma\delta}c_{\alpha\beta\gamma\delta}^{(2)}-b_{\alpha\beta}s^2 .
\end{equation}

\subsection{Magnetostriction near $T=T_c$ for amorphous dolids below the glass transition $T<T_g$}
 \label{MS}

 In our system, due to magnetization,  a compressive strain is generated along the $z$ direction. Further, the compressive stress obeys $\sigma_{zz} =0$ as the film has a free surface both in the nonmagnetic and magnetic states. Thus in the nonmagnetic state we can use Eq.~(\ref{Umechstrain3}) to write
\begin{equation}
\label{Umechstrain33}
\sigma_{zz} = 0 = c_{zz}^{(1)}  + \epsilon_{\gamma\delta} c_{zz\gamma\delta}^{(2)}  .
\end{equation}
 \noindent while in the magnetic state we can use Eq.~(\ref{Ustrain6}) to write
 \begin{equation}
\label{Ustrain6}
\sigma_{zz}(m) =  0 = c_{zz}^{(1)}+ \epsilon_{\gamma\delta}c_{zz\gamma\delta}^{(2)}+\gamma c_{zzzz}^{(2)}-b_{zz}s^2 .
\end{equation}
Subtracting Eq.~(\ref{Umechstrain33}) from Eq.~(\ref{Ustrain6}) we find the the magnetostriction coefficient
\begin{equation}
\label{magneto}
\gamma = \frac{b_{zz}s^2}{c_{zzzz}^{(2)}} =s^2\frac{\sum\limits_{i \neq j}\frac{1}{\langle r_{ij}\rangle }\frac{dJ(r_{ij})}{dr_{ij}}
(z_i-z_j)^2}
{\sum\limits_{i \neq j}
\Bigg[\frac{1}{r^2_{ij}} \frac{d^2 \phi}{d r^2_{ij}}-\frac{1}{r^3_{ij}}\frac{d\phi}{d r_{ij}}\Bigg] (z_i-z_j)^4}
\end{equation}
\noindent The first important result predicted by Eq.~(\ref{magneto}) is that the magnetostriction coefficient $\gamma$ scales quadratically with the magnetization. The second important result concerns the width dependence of $\gamma$ for a film of width $w_0$.
An accurate analysis of the results of the numerical simulations indicates that the saturation magnetostriction of a wide enough films depends on the width according to Eq.~(\ref{filmlamb}).
 We can use our theory to understand this dependence.
Using Eqs.~(\ref{Umechstrain4}) and ~(\ref{Uexstrain3}), we  can split formally the coefficient $b_{zz}$ in Eq.~(\ref{magneto}) into two contributions  $b_{zz}\approx b_{zz}^{\rm b} + b_{zz}^{\rm s}/w_0$
where the subscripts b and s stand for bulk and surface. Similarly we can write $c_{zzzz}^{(2)}\approx c_{zzzz}^{(2,\rm b)}+ c_{zzzz}^{(2,\rm s)}/w_0$. Thus for magnetic films of width $w_0$ we find
\begin{equation}
\label{magneto2}
\gamma (s,w_0) = s^2\frac{ w_0 b_{zz}^{\rm b} + b_{zz}^{\rm s}}{w_0 c_{zzzz}^{(2,\rm b)}+ c_{zzzz}^{(2,\rm s)}} \ .
\end{equation}
As $w_0$ tends to infinity we compute
\begin{equation}
\lambda_{\rm bulk} = \frac{b_{zz}^{\rm b}}{c_{zzzz}^{(2,\rm b)}} \ .
\end{equation}
In fact, we can use the general equation Eq.~(\ref{magneto}) to compute the magnetostriction coefficient of a film.
In contrast to films on a substrate considered in Section \ref{num} this approach corresponds to films with two free boundaries. Nevertheless, we can compare asymptotic bulk value of the saturated magnetostriction coeficient in these two cases.
Plotting the coefficients of $s^2$ in Eq.~(\ref{magneto}) multiplied by $w_0$ (in order to obtain the coefficient $b$) as a function of $w_0$ we find
 the red triangles in Fig.~\ref{fig9}. The best linear fit results in the estimate
\begin{equation}
\lambda_{\rm bulk} \approx 3.8 \times 10^{-3}\ .
\end{equation}
Comparing with the numerical result in Eq.~(\ref{numres}) we conclude that the
agreement between the mean-field theory and the simulations is very satisfactory.

Finally we can expand  Eq.~(\ref{magneto2}) in inverse powers of $w_0$ and the leading
result will read exactly like Eq.~(\ref{filmlamb}). The coefficient $\Lambda$ cannot be
directly compared between theory and experiment because it stems from two different
sources. One is purely geometric, particles in the center of the film have more
interaction than close to the two surfaces. The second comes form the interaction
between the film and substrate. In the theory we did not take particular care of
the interaction between the film and the substrate so the intercept in Fig.~\ref{fig9}.
\section{Summary and Discussion}
\label{summary}
In summary, we have presented a theory for the mechanical and magnetic responses of
amorphous solids which is equally applicable to a bulk sample or a film whose
width is in the nano scale. In this paper we focused on the magnetostriction as a good
measure of the interplay between mechanical strain and magnetic fields. Analytic theory
for all the other responses was offered both at $T=0$ or at finite temperature. We found that
for intermediate temperatures between $T_c$ and $T_g$ the nonaffine contribution to the
magnetostriction was negligible in the nano film. This simplifies the theoretical calculation
of the magnetostriction coefficient which is found to be in good agreement with the numerical
simulations.
Interestingly enough, both in bulk and in nano film the magnetostriction coefficient is proportional
to $s^2$ and therefore to the square of the magnetization. It is expected that at low
temperatures, $0<T\ll T_c$ the non-affine contribution should be significant, since at $T=0$
it is of the same order as the Born contribution. It is therefore interesting to examine this
issue both in experiments and in simulations at this range of temperatures.

\acknowledgments
This work had been supported in part by the US-Israel Bi-national Foundation and by
the Israel Science Foundation.

\appendix
\section{Monte Carlo Eqilibration}
\label{A}
The Monte Carlo (MC) simulations were performed in both NPT (in a bulk phase) and NVT (in a  film) ensembles.
In a bulk phase we start with initial face-centered-cubic arrangement of $A$ type particles with periodic boundary conditions in three directions. Then randomly chosen $20 \%$   of the particles were changed to $B$ type.
The initial configuration of a  film (of desired height) was cut along the z axis from a  bulk glass  and put on top of the crystalline substrate.  Then the system (the bulk phase or the film) was equilibrated at high temperature $T=5$.
Once equilibrated, the systems were quenched instantaneously to $T=0.23$. In each NVT ensemble MC-sweep we attempt to move each particle once. We chose the maximum position displacement such that the acceptance ratio of the trial moves was around $30\%$.
In the case of NPT ensembles, in addition to the trial moves we attempt to change the length of the simulation box in every $20$ MC-sweeps. We chose the maximum change in box-length such that the acceptance ratio of the trial moves was around $30\%$. Optimum particle
displacements and changes in the box-length are obtained for $200000$ MC-sweeps  before starting to gather thermal statistics. To update the
spins, we use Wolf's cluster algorithm \cite{Wolf89} when there is no external magnetic field. We made two modifications to this algorithm. Firstly, concentrating on any given particle $i$ we refer to
its neighbors, as any particle $j$ that resides within a distance
of $2.5$ from it. Secondly, the coupling defined by $J(r_{ij})$ (see Eq.~(\ref{Yuk})) is not a constant as in a common lattice problem. We attempt the particle move and spin flip in the following sequence: two sweeps, in each of which we
attempt to
move each particle once, are followed by the construction of one Wolf cluster after which the Monte Carlo proceeds
with the next two sweeps. In the presence of magnetic field the Wolf's cluster algorithm is not effective. Hence we apply single spin flip algorithm
in which we attempt to randomly flip each spin once.

\section{Responses at zero temperature}
\label{zero}
 All the important response functions exhibited by magnetic amorphous solids at $T=0$ have been studied in great detail and can be expressed in terms of the eigenvalues and eigenfunctions of a  Hessian matrix $\B H$  for $N$ particles in $d$ dimensions where
\begin{eqnarray}
\label{hessian}
\B H^{(\B r \B r)}_{ij} & \equiv  & \frac{\partial^2U}{\partial \B r_i\partial \B r_j}  \qquad  \text{($dN \times dN$ matrix)}\nonumber \\
\B H^{(\B r \B S )}_{ij} & \equiv  & \frac{\partial^2U}{\partial \B r_i\partial \B S_j} \qquad  \text{($dN \times dN$ matrix)}\nonumber \\
\B H^{(\B S \B r)}_{ij} & \equiv  & \frac{\partial^2U}{\partial \B S_i\partial \B r_j} \qquad  \text{($dN \times dN$ matrix)}\nonumber \\
\B H^{(\B S \B S)}_{ij} & \equiv  & \frac{\partial^2U}{\partial \B S_i\partial \B S_j} \qquad  \text{($dN \times dN$ matrix)} ,
\end{eqnarray}
and four  `mismatch forces'  $\B \Xi$ that represent the forces and torques on the particles before the non-affine flows ensure new local minima for the particle positions and spins.  For notational simplicity let us assume that we can replace the stress tensor $\epsilon_{\alpha\beta}$ by a scalar $\gamma$ here, then the particle positions and spins can be written  $\{\B r_i(\gamma, B )\}, \{{\B S}_i(\gamma, B)\}$.  Then the mismatch forces can be written
\begin{eqnarray}
\label{mismatch}
\B \Xi^{(\gamma, \B r )}_i & \equiv  & \frac{\partial^2U}{\partial \gamma \partial \B r}_i \nonumber \\
\B \Xi^{(\gamma, \B S )}_i & \equiv  & \frac{\partial^2U}{\partial \gamma \partial \B S}_i \nonumber \\
\B \Xi^{(B, \B r)}_i & \equiv  & \frac{\partial^2U}{\partial B \partial \B r}_i \nonumber \\
\B \Xi^{(B, \B S )}_i & \equiv  & \frac{\partial^2U}{\partial B \partial \B S}_i
\end{eqnarray}
\noindent In terms of these Hessian and mismatch forces we can find all the modulii that describe the mechanical and magnetic properties of magnetic glasses.\\
\noindent Thus the shear modulus takes the form
\begin{equation}
\label{shearmodulus2}
\mu (\gamma, B)  =  \frac{\partial^2 U }{\partial \bf \gamma^2}{\bf |}_{B}  - {\B \Xi^{(\bf \gamma )}} \cdot \B H^{-1}\cdot \B \Xi^{(\gamma )}
\end{equation}
Note the non-affine contribution reduces the shear modulus.

\noindent The magnetic susceptibility $\chi (\gamma ,B)$ can similarly be expressed in terms of a classic thermodynamic form that exists for crystalline solids and an additional term required for magnetic equilibrium in the case of amorphous solids
\begin{equation}
\label{susceptibility}
\chi (\gamma, B) = -\frac{\partial^2 U }{\partial B^2}{\bf |}_\gamma  + {\B \Xi^{(B)}} \cdot \B H^{-1}\cdot \B \Xi^{(B)}
\end{equation}
\noindent Here the additional positive definite form exists  due to the existence of nonaffine flows that can help minimize the potential energy of the magnetic glass.
\noindent Magnetostriction can be measured from the change of stress of a specimen with changing magnetic field B
\begin{equation}
\label{magnetostriction}
\chi_{\sigma,B}(\gamma, B)   =  \frac{d \sigma }{d B} {\bf |}_\gamma  = \frac{\partial^2 U }{\partial B \partial \gamma} - {\B \Xi^{(\gamma)}} \cdot \B H^{-1}\cdot \B \Xi^{(B)}
\end{equation}
\noindent While magneto elasticity and magneto plasticity involve the magnetic response of a material to applied strain
\begin{equation}
\label{magnetoplasticity}
\chi_{M,\gamma} (\gamma, B)   =  \frac{d M }{d \gamma} {\bf |}_{B}   = -\frac{\partial^2 U }{\partial \gamma \partial B} + {\B \Xi^{(B)}} \cdot \B H^{-1}\cdot \B \Xi^{(\gamma)} .
\end{equation}

\noindent Note that  for crystalline solids we have the Maxwell relation between magnetostriction and magneto elasticity $\frac{\partial \sigma }{\partial B} {\bf |}_\gamma = - \frac{\partial M }{\partial \gamma} {\bf |}_{B}$. For metallic glasses as the Hessian matrix is hermitian  we  have an analogous Maxwell relationship
\begin{equation}
\label{maxwell}
\frac{d\sigma }{d B} {\bf |}_\gamma = - \frac{d M }{d\gamma} {\bf |}_{B} \nonumber
\end{equation}

\end{document}